\documentclass[prc,preprint,aps,groupedaddress,showpacs]{revtex4}

\usepackage{graphicx}
\begin{document}

\title{Lifetime Measurements in $^{182,186}$Pt} 

\author{J. C. Walpe}
\thanks{Present address: ADESA, Inc., Carmel, IN 46032.} 
\author{U. Garg}
\author{S. Naguleswaran}
\thanks{Present address: Center for Defence Communication and Information
Networking, The University of Adelaide, Adelaide SA5005, Australia.} 
\author{J. Wei}

\affiliation{Department of Physics, University of Notre Dame, Notre Dame, IN
46556}

\author{W. Reviol}
\affiliation{Department of Chemistry, Washington University,
St. Louis, MO 63130} 

\author{I. Ahmad}
\author{M. P. Carpenter}
%\author{R. V. F. Janssens}
\author{T. L. Khoo}
\affiliation{Physics Division, Argonne National Laboratory, Argonne, IL 60439}

\date{\today}

\begin{abstract}
Lifetimes in the yrast bands of the nuclei $^{182,186}$Pt have been measured using 
the Doppler-shift Recoil Distance technique. The results in both cases 
%provide evidence for shape coexistence in these nuclei: 
{\em viz.} a sharp increase in B(E2) values at very low spins,
may be interpreted as resulting from a mixing between two bands of  
different quadrupole deformations.  
\end{abstract}


\maketitle

Shape-coexistence 
has been observed in many nuclei in the $A \sim 180$ region \cite{wood1}.
For example, in the $^{178-188}$Hg (Z=80) nuclei, bands of 
states of spin-parity 0$_{1}^{+}$, 2$_{1}^{+}$, 4$_{1}^{+}$, ... and 0$_{2}^{+}$, 2$_{2}^{+}$, 
4$_{2}^{+}$, ..., have been observed with similar intensities \cite{ham,carp2} and are known (via 
energy-spacing and B(E2) measurements) to have different quadrupole 
deformations \cite{cole,bera}; the lower-energy bands exhibit oblate shape 
characteristics, and the higher-energy bands exhibit prolate 
characteristics \cite{beng}. Indeed, all mean-field calculations \cite{aberg, rod1,rod2,rod3} predict a 
coexistence of prolate and oblate shapes in this
region of nuclei. There is also evidence to support shape coexistence 
in the light Pt (Z=78) nuclei \cite{garg,drac,drac2}.
In particular, the B(E2) values in the yrast band of $^{184}$Pt undergo 
a dramatic increase in going from the 2$^{+}$ state to the 6$^{+}$ state and 
this increase has been interpreted as resulting from a mixing, at low spins, 
between two bands with deformations of different magnitude \cite{garg}. An extensive
discussion of the deformation-driving aspects of all the quasiparticles, as well
as shape-coexistence, in $^{184}$Pt has been provided in Ref.\cite{mpc}.

The shape-coexistence picture in the Pt nuclei has engendered extensive
theoretical 
interest in recent years \cite{libby1,libby2,pvi,heyde1,heyde2}. In
particular, there has been a controversy regarding
whether shape evolution in the Pt isotopes involves
intruder states \cite{pvi,heyde1,heyde2}, or can be understood
without invoking the intruder states \cite{libby1,libby2}. Most recently,
Garc\'{i}a-Ramos {\em et al.} \cite{heyde2} have established that
configuration-mixing is an essential feature of these 
 nuclei, although it is not apparent when
considering only a limited set of data; they termed this ``concealed
configuration
mixing'' and have suggested further spectroscopic measurements where this
mixing might be clearly revealed. Lifetime measurements, and the extracted
transition probabilities, up to the states beyond the mixing region provide
such additional spectroscopic information.

In this Brief Report, we present results from lifetime measurements in the 
$^{182,186}$Pt nuclei, using the Doppler-shift recoil distance method
(RDM). 
The primary aim of these measurements was to extend the $^{184}$Pt results to the nearby Pt isotopes in order to
develop a better understanding of the nuclear structure properties in this transitional region. No previous lifetime information was
available on the excited states in these two nuclei, save that of the
first 2$^+$ state in $^{186}$Pt, obtained from decay of $^{186}$Au
\cite{finger}.

Two separate RDM experiments were carried out at the ATLAS facility at the 
Argonne National Laboratory,
using the $^{154}$Sm($^{36}$S, 4n)$^{186}$Pt and $^{122}$Sn($^{64}$Ni, 4n)$^{182}$Pt
reactions
at beam energies of 167 MeV and 295 MeV, respectively. Statistical model
calculations (with the code CASCADE) and brief 
excitation function measurements were employed to determine the beam energies 
for the optimal population of the $4n$ channels. In the case of $^{186}$Pt,
the target was enriched $^{154}$Sm evaporated, to a thickness of
500 $\mu g/cm^{2}$, onto a stretched 1.2 mg/cm$^{2}$-thick Au foil. For the
$^{182}$Pt experiment, 
the target used was similarly prepared with enriched $^{122}$Sn evaporated onto a 
1.5 mg/cm$^{2}$-thick Au foil to a thickness of 850 $\mu g/cm^{2}$. The targets
were 
mounted in the Notre Dame ``plunger'' device, with the Au surface facing the
beam. The plunger device consisted of three dc actuators 
which are used for precision placement of the target foil with respect to a 
fixed stopper foil (a stretched, self-supporting, Au foil of 10.0 mg/cm$^{2}$ 
thickness in the case of $^{186}$Pt and
25 mg/cm$^{2}$ thickness in the case of $^{182}$Pt).
Gamma-ray spectra were recorded
using the Argonne-Notre Dame $\gamma$-ray Facility, which consisted
of twelve Compton-suppressed HPGe detectors and a 50-element BGO multiplicity
array. Four detectors each were placed at 
angles of 34$^{\circ}$ (``forward-angle''), 90$^{\circ}$, and 146$^{\circ}$
(``backward-angle'') with respect to the 
beam direction. Data were collected for approximately 3 hours at each of 18 and 
24 target-to-stopper distances for $^{186}$Pt and $^{182}$Pt, respectively, 
ranging from the closest attainable distances (corresponding to electrical 
contact) of 15 $\mu m$ ($^{186}$Pt)
and 9 $\mu m$ ($^{182}$Pt) to a maximum distance of $\sim$10000 $\mu m$ in each 
case. This resulted in an effective measurable lifetime range of $\sim$2 ps to 
$\geq$1 ns.

\begin{figure}[b]
\includegraphics[scale=2.0]{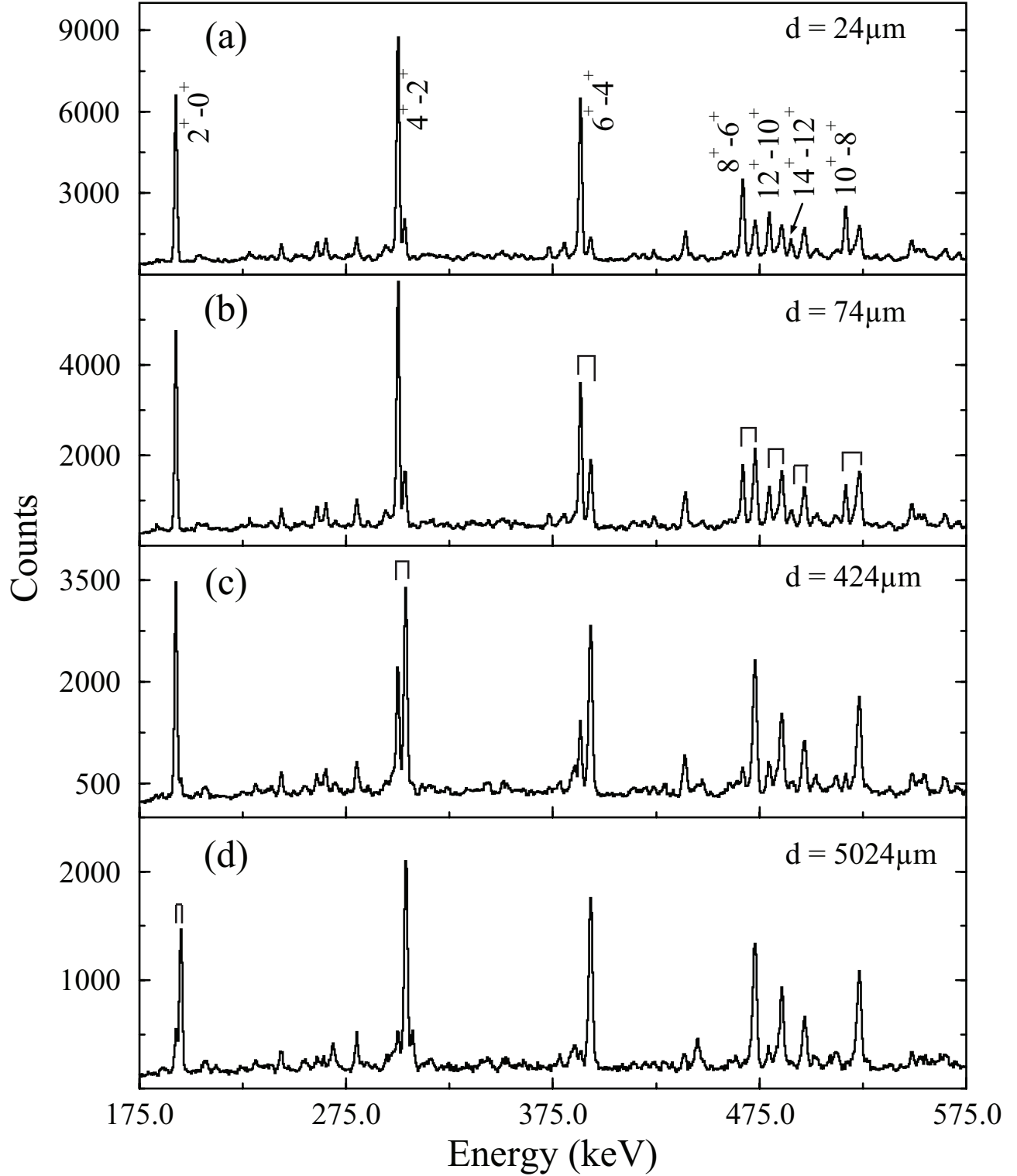}
%\vspace*{8cm}
\caption[]{ Sample spectra from the ``forward-angle'' detectors for $^{186}$Pt at the
indicated recoil distances. The transitions of interest are marked. The relative
increase in the intensities of the ``shifted'' peaks with increasing distance 
is clearly visible.}
\label{spectra1}
\end{figure}

\begin{figure}
\includegraphics[scale=2.0]{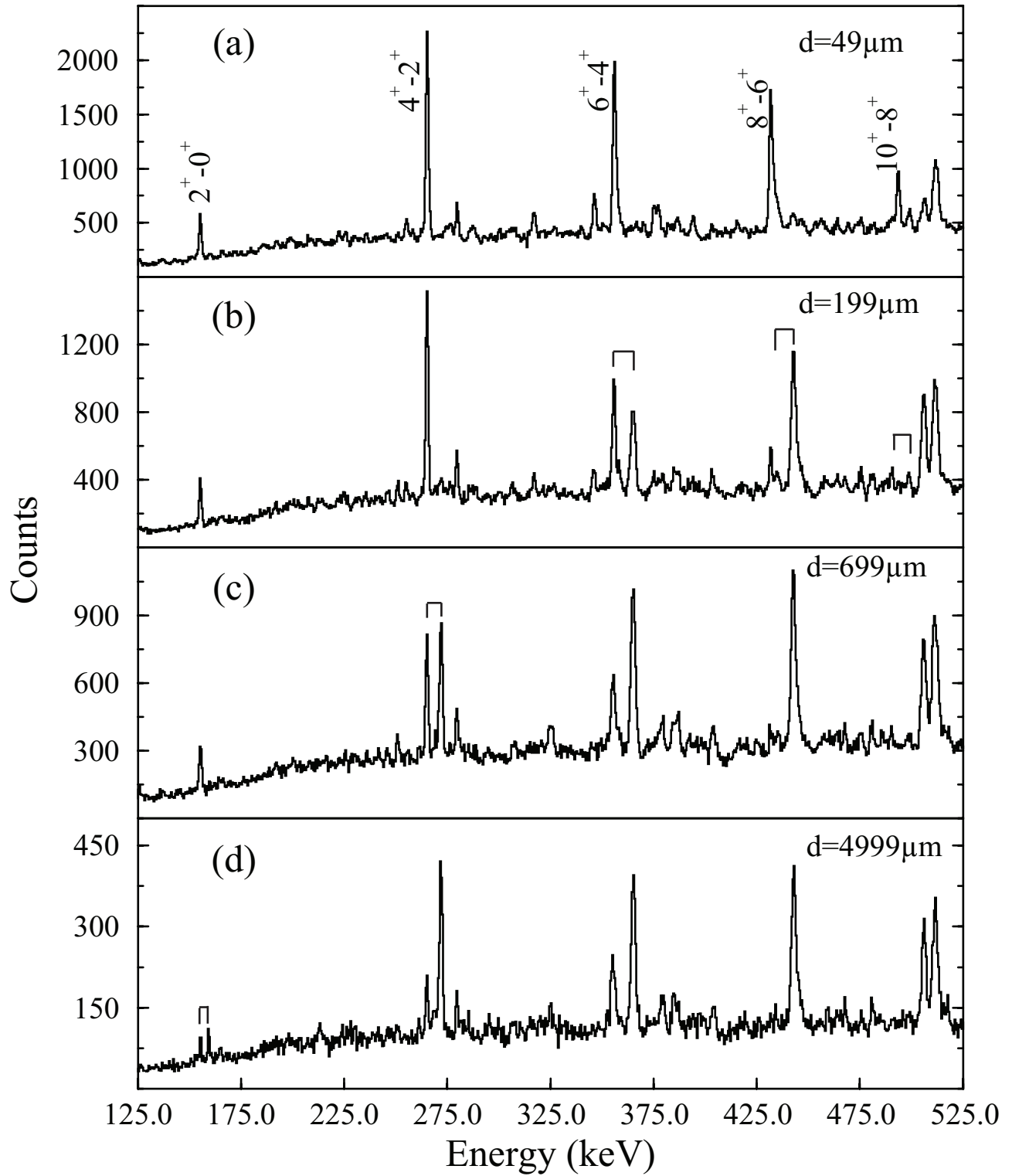}
%\vspace*{8cm}
\caption[]{ Same as Fig. 1, but for $^{182}$Pt.}
\label{spectra2}
\end{figure}

Sample spectra for several recoil distances, taken with the detectors placed
at 34$^{\circ}$, are shown in Fig.~\ref{spectra1} and Fig.~\ref{spectra2}, with 
the corresponding recoil distance given in the upper right corner of each 
spectrum. 
The level schemes of these nuclei are known from Refs.\cite{lev1,lev2} and 
the transitions of interest are labelled.
Lifetime information was reliably extracted for 
yrast transitions up to the 16$^{+}$ (14$^{+}$) level in $^{186}$Pt ($^{182}$Pt). For each transition,
 a set of ratios, $R_{d}$, defined as the ratio of the intensity of the
unshifted $\gamma$-ray peak
to the total intensity of the unshifted and the associated shifted peak at
recoil distance $d$, were determined.
Each set of $R_{d}$ values defined a decay curve for a given transition. All such
$R_{d}$ curves were fitted simultaneously with a combination of exponential functions and
the lifetime for each level was extracted from these fits. The computer code 
LIFETIME \cite{wells} was employed in the fitting procedure. 
This code allows for all the ``standard'' corrections to be applied to the 
data: changes in the solid angle subtended by the detectors due to the changing ion position 
along the flight path; changes in solid angle subtended by the detector due to the 
relativistic motion of the ion; changes in the angular distribution due to the
attenuation of alignment while the ion was in flight; and, slowing of the
ion in the stopper material. In addition, corrections were made to the data to
account for the detector efficiencies and internal conversion.  
A crucial aspect of data analysis was accounting for the effects of cascade
feeding, both observed and unobserved, from higher-lying states. A two-path
feeding process into the level was assumed for each instance where 
the observed intensity feeding into the level was less than the observed
intensity decaying out of it.  One of the feedings is from the
next highest transition in the yrast cascade, which represents the cumulative effect of the decay of all higher members of the cascade, while the other represents
all unobserved feeding, {\em i.e.}, feeding from the $\gamma$-ray
continuum and non-yrast states.  The relative intensities for the observed
states were determined from the data collected by the detectors placed at
90$^{\circ}$.  Initial relative intensities for the levels representing
the unobserved feeding were determined by taking the difference between the
observed intensity into a given level and the observed intensity out of the
said level. 

\begin{figure}
\includegraphics[scale=0.55]{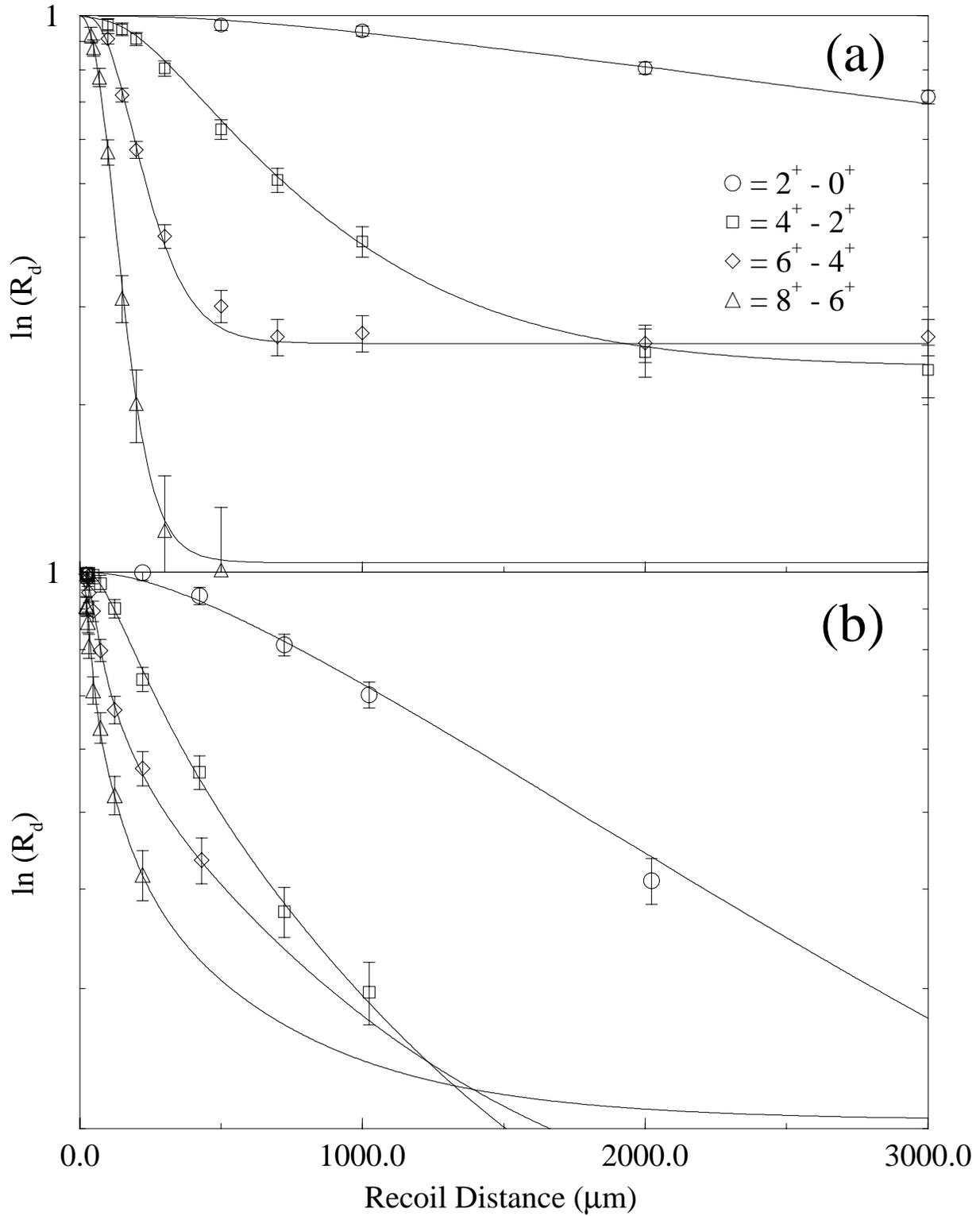}
\vspace*{6.5cm}
\caption[]{ Fits to the ratios R$_{d}$ as a function of the recoil distance
for the lowest yrast transitions in (a) $^{186}$Pt and (b) $^{182}$Pt, as 
obtained from the code LIFETIME.}
\label{fits}
\end{figure}

The resulting fits to the experimental data are presented in Fig.~\ref{fits} and 
the extracted lifetimes as well as the corresponding B(E2) values are given 
in Tables I and II; preliminary results have been reported previously
\cite{wei,walpe}. We note that the lifetime obtained for the 2$^+$ state in
$^{186}$Pt is consistent with the ``adopted value'' in the compilation by
Raman {\em et al.} \cite{raman,finger}. Further, of all the theoretical
predictions listed in Ref.\cite{raman}, the B(E2)'s for
the 2$^+$ states in $^{182,184,186}$Pt obtained in our measurements are
closest to those from the Woods-Saxon Model (WSM in Ref.\cite{raman}). 

As can be seen in the Tables and Fig. 4, there is indeed a sharp increase in
the B(E2) value when going from the 2$^{+}$ to the 6$^{+}$ state in each case, 
similar to the observation in $^{184}$Pt \cite{garg}. 
In a deformed rotor interpretation, this change in B(E2) value corresponds to 
an increase in deformation
from $\beta_{2}$=0.20 for the 2$^{+}$ state to $\beta_{2}$=0.25 for the 6$^{+}$ 
state in $^{186}$Pt and from $\beta_{2}$=0.19 for the 2$^{+}$ state to 
$\beta_{2}$=0.25 for the 6$^{+}$ state in $^{182}$Pt. [If this were an
undisturbed rotational band, the $\beta_{2}$ value would be nearly a constant.].
These results are in qualitative agreement with the calculations presented in Ref.\cite{mpc}.

\begin{figure}
\includegraphics[scale=1.7]{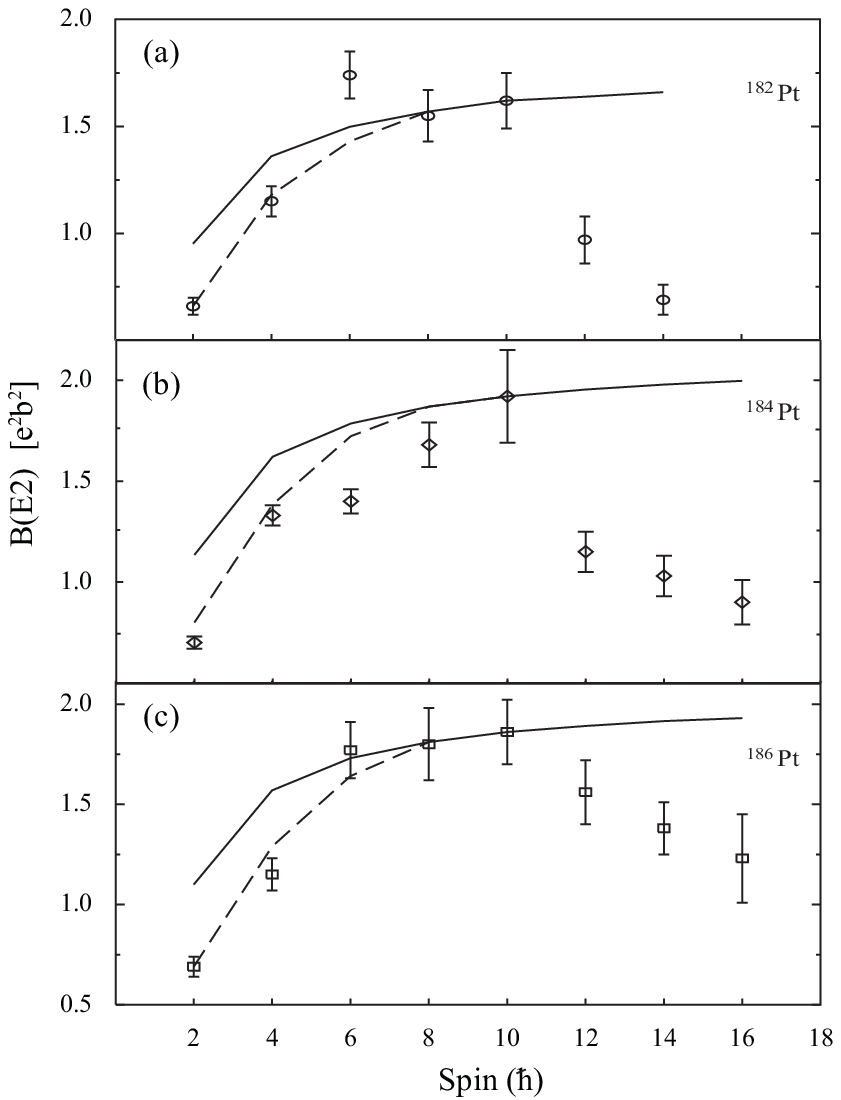}
%\vspace*{-2cm}
\caption[]{ B(E2) vs. the spin of the depopulating state for $^{182}$Pt
(panel (a)) and $^{186}$Pt (panel (c)).  The experimental results from the
present work are shown, along with the 
expected results for a rigid rotor with deformation corresponding to the
10$^{+} \rightarrow $8$^{+}$ transition (solid lines) and the results of
two-band mixing calculations (dashed lines; see text). The results for
$^{184}$Pt (taken from Ref. \cite{garg}) are also shown (panel (b)) for comparison.}
\label{be2vspin}
\end{figure}

A likely explanation for this increase in the associated 
deformation is that there is a 
mixing of states at low spins. Dracoulis {\em et al.} \cite{drac,drac2} had 
interpreted such increase in the deformation in the Pt nuclei as resulting from 
a mixing of coexisting bands and, in performing two-band mixing
calculations in $^{176-178}$Pt, obtained results consistent with the observed behavior
of the bandhead energies and mixing matrix elements. We have performed
similar band-mixing calculations for the nuclei studied here, the results of 
which are 
shown in Fig.~\ref{be2vspin}; the results
from $^{184}$Pt \cite{garg} are also included for comparison. It has been
assumed in these calculations that the bands mix at low spins only, and that 
the 8$^{+}$ level in both cases is completely unmixed.  The unperturbed energies of 
the yrast bands (called B1) were then calculated using a rotational constant of 
$\hbar^{2}/2{\cal J}$ = 15.51 (14.36) keV for $^{186}$Pt ($^{182}$Pt),
obtained from the respective $8^{+} \rightarrow 6^{+}$ transition energies.
The mixing bands (called B2) were assumed to be 
built on the low-lying 0$^{+}_{2}$ states, at 472 (500) keV in $^{186}$Pt 
($^{182}$Pt) \cite{hebbing}.  
Calculations for the mixing between the 0$_{1}^{+}$ state (in B1) and the 
0$_{2}^{+}$ state (in B2) resulted in interactions, V$_{int}$, between the unmixed states of 259 (237)
keV for $^{186}$Pt ($^{182}$Pt). These mixing interactions are comparable to
those obtained in two-band model calculations by Thiamo\'{v}a and Van Isacker
\cite{thia}. The calculations also imply that the 
ground-state band in $^{186}$Pt is composed of 56$\%$ B1 (larger 
deformation) and 44$\%$ B2 (smaller deformation), whereas the ground-state band 
in $^{182}$Pt is composed of 66$\%$ B1 and 34$\%$ B2.  These results are 
very similar to those obtained in Ref. \cite{garg} for $^{184}$Pt and are consistent with the 
theoretical results in Refs. \cite{heyde1,heyde2}, supporting a band-mixing interpretation of this B(E2) increase at low spins in these nuclei.
%They are also in agreement with 
%the prediction that the more-deformed band (B1) lies lower in energy than the 
%lesser-deformed band (B2) \cite{wood1, garg}, which is in contrast to the
%observed behavior in the Hg nuclei, where the spherical or slighly oblate band is seen to
%lie lower in energy than the prolate-deformed sequence \cite{ham2}.

The results presented in Fig.~\ref{be2vspin} also exhibit a steep decline in
B(E2) values beyond the 10$^{+}$ state in both nuclei. Some of this decline
comes naturally because of the alignment of the $i_{13/2}$ neutrons
that is associated with a loss of collectivity in a limited spin range.
In many transitional nuclei, a similar decline in B(E2)'s has been attributed to
the possible onset of triaxiality, with positive values
of $\gamma$, the classic example being that of the Yb nuclei \cite{john}. 
%Indeed, Ma {\em et al.} \cite {ma} had argued quite some time ago that 
%the alignment of $i_{13/2}$ neutrons drives the nucleus to a
%triaxial shape. 
This reduction in B(E2)'s (and, hence, in $Q_t$) is also consistent with the theoretical results
presented in Ref. \cite{rod2}.

Very recently, results of another lifetime measurement on $^{182}$Pt
have become available \cite{182pt}. In that work,
lifetimes have been measured up to the 10$^+$ state and
their results are in good agreement with those presented here,
albeit the quoted uncertainties
in Ref. \cite{182pt} are larger. They also have presented calculations within
the Interacting Boson Model and the General Collective Model, both of which
indicate shape coexistence in this nucleus, in general agreement with the
conclusions presented here.

In summary, we have measured the lifetimes of the yrast states in $^{182}$Pt and
$^{186}$Pt up to spins 14$^{+}$ and 16$^{+}$, respectively, using the 
Doppler-shift recoil distance technique.  A sharp 
increase in the B(E2) value in going from the 2$^{+}$ state to the 6$^{+}$ 
state is observed in both nuclei. This increase can be understood, in the
traditional way, in terms of
the mixing of coexisting bands of different deformations at low spins and 
two-band mixing calculations are in good 
agreement with the observed experimental results. We wish to emphasize, however, that
the two-band mixing interpretation of these data might not be unique. Indeed, it would appear that the 
observed B(E2)'s in $^{182}$Pt can be reproduced reasonably well in IBA-1 calculations
using the parameters provided in Ref. \cite{libby2}. Further measurements, especially of lifetimes of
non-yrast states, could be very useful in unambiguously deciding between the competing theoretical
approaches to understanding the nuclear structure in this region.

We wish to acknowledge K.B. Beard (Notre Dame), I. Bearden (Purdue Univ.), R.V.F. Janssens (Argonne),
S. Shastry (SUNY, Plattsburgh), and D. Ye (Notre Dame) for their important
contributions to these measurements. This work has been supported in part by the
National Science Foundation (Grant No. PHY-1068192) and the U.S. Department of
Energy, Nuclear Physics Division, under contract no. DE-AC02-06CH11357.

\newpage
\begin{table}
\caption{ Lifetimes, associated B(E2) values, and the extracted transition quadrupole moments, $Q_t$, for $^{186}$Pt}
\
\begin{tabular}{ccccc}
\colrule
E$_{\gamma}$(keV)&$I_{i} \rightarrow I_{f}$&$ \tau$(ps)&B(E2)$ \downarrow$(e$^{2}b^{2}$)&Q$_{t}$
($eb$)\\
\colrule
191.4&$2^{+} \rightarrow 0^{+}$&318$\pm$24&0.69$\pm$0.05&5.87$\pm$0.22 \\
298.8&$4^{+} \rightarrow 2^{+}$&27.3$\pm$1.9&1.15$\pm$0.08&6.37$\pm$0.22 \\
387.2&$6^{+} \rightarrow 4^{+}$&5.1$\pm$0.4&1.77$\pm$0.14&7.53$\pm$0.30 \\
465.4&$8^{+} \rightarrow 6^{+}$&2.0$\pm$0.2&1.80$\pm$0.18&7.42$\pm$0.37 \\
515.4&$10^{+} \rightarrow 8^{+}$&1.2$\pm$0.1&1.86$\pm$0.16&7.43$\pm$0.31 \\
478.8&$12^{+} \rightarrow 10^{+}$&2.0$\pm$0.2&1.56$\pm$0.16&6.76$\pm$0.34 \\
489.2&$14^{+} \rightarrow 12^{+}$&2.1$\pm$0.2&1.38$\pm$0.13&6.30$\pm$0.30 \\
571.1&$16^{+} \rightarrow 14^{+}$&1.1$\pm$0.2&1.23$\pm$0.22&5.93$\pm$0.54 \\
658.0&$18^{+} \rightarrow 16^{+}$&1.6$\pm$0.2\footnotemark[1]&& \\
\colrule
\end{tabular}
\vspace{0.5cm}
\caption{ Lifetimes, associated B(E2) values, and the extracted transition quadrupole moments, $Q_t$, for $^{182}$Pt}
\
\begin{tabular}{ccccc}
\colrule
E$_{\gamma}$(keV)&$I_{i} \rightarrow I_{f}$&$ \tau$(ps)&B(E2)$ \downarrow$(e$^{2}b^{2}$)&Q$_{t}$
($eb$)\\
\colrule
155.0&$2^{+} \rightarrow 0^{+}$&709$\pm$43&0.66$\pm$0.04&5.77$\pm$0.17 \\
264.3&$4^{+} \rightarrow 2^{+}$&47.5$\pm$2.9&1.15$\pm$0.07&6.37$\pm$0.19 \\
355.0&$6^{+} \rightarrow 4^{+}$&7.8$\pm$0.5&1.74$\pm$0.11&7.45$\pm$0.24 \\
430.9&$8^{+} \rightarrow 6^{+}$&3.4$\pm$0.3&1.55$\pm$0.12&6.88$\pm$0.27 \\
492.6&$10^{+} \rightarrow 8^{+}$&1.7$\pm$0.2&1.63$\pm$0.13&6.95$\pm$0.28 \\
543.5&$12^{+} \rightarrow 10^{+}$&1.7$\pm$0.1&0.97$\pm$0.11&5.33$\pm$0.30 \\
590.0&$14^{+} \rightarrow 12^{+}$&1.6$\pm$0.2&0.69$\pm$0.07&4.47$\pm$0.23 \\
628.5&$16^{+} \rightarrow 14^{+}$&3.3$\pm$0.4\footnotemark[1]&& \\
\colrule
\vspace{0.25cm}
\footnotetext[1]{ Lifetimes of the 18$^{+}$ and 16$^{+}$ levels, respectively, could 
not be separated from the side feeding lifetimes.  The values given are therefore upper 
limits.}
\end{tabular}
\end{table}

\newpage
\newpage
\newpage
\newpage


\begin{thebibliography}{99}
\bibitem{wood1} Kris Heyde and John L. Wood, Rev Mod. Phys. {\bf 83}, 1467 (2011).
\bibitem{ham} J.H. Hamilton, P.G. Hansen, and E.F. Zganjar, Rep. Prog Phys.
{\bf 48}, 631 (1985).
\bibitem{carp2} M.P. Carpenter {\em et al.}, Phys. Rev. Lett. {\bf 78}, 3650 (1997).
\bibitem{cole} J.D. Cole {\em et al.}, Phys. Rev. C {\bf 30}, 1267 (1984).
\bibitem{bera} R. Beraud {\em et al.}, Nucl. Phys. {\bf A284}, 221 (1977).
\bibitem{beng} R. Bengtsson, T. Bengtsson, J. Dudek, G. Leander, W. Nazarewicz, 
and J.-Y. Zhang, Phys. Lett. {\bf B183}, 1 (1987).
\bibitem{aberg} S. Aberg, H. Flocard, and W. Nazarewicz, Annu. Rev. Nucl. Part.
Sci. {\bf 40}, 439 (1990).
\bibitem{rod1} P. Sarrigueren {\em et al.}, Phys. Rev. C {\bf 77}, 064322 (2008).
\bibitem{rod2} R. Rodr{\'i}guez-Guzm{\'a}n  {\em et al.}, Phys. Rev. C {\bf 81}, 024310 (2010).
\bibitem{rod3} K. Nomura {\em et al.}, Phys. Rev. C {\bf 83}, 014309 (2011).
\bibitem{garg} U. Garg {\em et al.}, Phys. Lett. {\bf B180}, 319 (1986).
\bibitem{drac} G. Dracoulis, A.E. Stuchbery, A.P. Byrne, A.R. Poletti, S.J.
Poletti, J. Gerl, and R.A. Bark,  J. Phys. G {\bf 12}, L97 (1986).
\bibitem{drac2} G.D. Dracoulis, Phys. Rev. C {\bf 49}, 3324 (1994).
\bibitem{mpc} M.P. Carpenter {\em et al.}, Nucl. Phys. {\bf A513}, 125 (1990).
\bibitem{libby1} E.A. McCutchan, R.F. Casten, and N.V. Zamfir, Phys. Rev. C
{\bf 71}, 061301(R) (2005).
\bibitem{libby2} E.A. McCutchan, and N.V. Zamfir, Phys. Rev. C
{\bf 71}, 054306 (2005).
\bibitem{pvi} Irving O. Morales {\em et al.}, Phys. Rev. C {\bf 78},
024303 (2008).
\bibitem{heyde1} J.E. Garc\'{i}a-Ramos and K. Heyde, Nucl. Phys. {\bf A825}, 39
(2009).
\bibitem{heyde2} J.E. Garc\'{i}a-Ramos, V. Hellemans, and K. Heyde , Phys. Rev. C
{\bf 84}, 014331 (2011).
\bibitem{finger} M. Finger {\em et al.}, Nucl. Phys. {\bf A188}, 369 (1972).
\bibitem{lev1} D.G. Popescu {\em et al.}, Phys. Rev. C {\bf 55}, 1175 (1997).
\bibitem{lev2} C.M. Baglin, Nucl. Data Sheets {\bf 82}, 1 (1997).
\bibitem{wells} Program LIFETIME, obtained from Prof. J.C. Wells, Tennessee 
Technological University, Cookville, TN.
\bibitem{wei} J. Wei {\em et al.}, Bull. Am. Phys. Soc. {\bf 35},
1016 H6 7 (1990).
\bibitem{walpe} J.C. Walpe {\em et al.}, Bull. Am. Phys. Soc. {\bf 39},
1419 DD4 (1994).
\bibitem{raman} S. Raman, C.W. Nestor, Jr., and P. Tikkanen, At. Data Nuc.
Data Tables {\bf 78}, 1 (2001).
\bibitem{hebbing} G. Hebbinghaus {\em et al.}, Z. Phys. {\bf A328}, 387 (1987).
\bibitem{thia} G. Thiamo\'{v}a and P. Van Isacker, Phys. Scr. {\bf 64}, 23 (2001).
%\bibitem{ham2} J.H. Hamilton {\em et al.}, Phys. Rev. Lett. {\bf 35}, 
%562 (1975).
\bibitem{john} N.R. Johnson, Prog. Part. Nuc. Phys. {\bf 28}, 215 (1992).
%\bibitem{ma} W.C. Ma {\em et al.}, Phys. Rev. C {\bf 47}, R5 (1993).
\bibitem{182pt} K.A. Gladinski {\em et al.}, Nucl. Phys. {\bf A877}, 19 (2012).

\end{thebibliography}
\end{document}